\documentclass{article}
\usepackage[T1]{fontenc}
\usepackage{epsfig,calc,amsfonts,natbib}
\usepackage{epstopdf}
\long\def\comment#1{}
\def\kurt{\kappa}
\def\skew{{\cal{S}}}

\def\Re{{\mathbb R}}
\def\tabb{\phantom{Hey.}}

\def\bbox#1{\vbox{\hbox{\includegraphics[width=\textwidth*\real{0.8}]{#1}}}}

\begin{document}
\author{Ben Klemens\\US Census Bureau\\
    ben.klemens@census.gov.\footnote{This paper originated as work done at Caltech, under
        the guidance of Matt Jackson, Kim Border, and Peter Bossaerts.
        The agent-based modeling work was done at the
        Brookings Institution's Center for Social and Economic Dynamics,
    and the author thanks the CSED's members for their support. Thanks also to Josh Tokle
    and Taniecea Arceneaux of the United States Census Bureau.}}
\title{A Peer-based Model of Fat-tailed Outcomes}
\maketitle
\begin{abstract}
It is well known that the distribution of returns from various financial instruments are
leptokurtic, meaning that the distributions have ``fatter tails'' than a
Normal distribution, and have skew toward zero. This paper presents a graceful micro-level
explanation for such fat-tailed outcomes, using agents whose private valuations have
Normally-distributed errors, but whose utility function includes a term
for the percentage of others who also buy. 
\end{abstract}

\section{Introduction}

Many researchers have pointed out that day-to-day returns on equities have ``fat tails,''
in the sense that extreme events happen much more frequently than would be predicted
by a Normal distribution, and have skew toward zero, meaning that extreme negative
returns are more likely than extreme positive returns. This has been re-verified by
many of the sources listed below. The fat tails of actual equity return distributions is
far from academic trivia: if extreme events are more likely than predicted by
a Normal distribution, models based on Normally-distributed returns can systematically
under-predict risk.

Here, I present an explanation for the non-Normality of equity returns using a
micro-level model where agents observe and emulate the behavior of others.  There are
several reasons for rational agents to take note of the actions of other rational
agents; the model here is agnostic as to which best describes real-world agents,
but given some motivation to emulate others, I show that the
wider-than-Normal distribution of equity returns follows.

From the tulip bubble of 1637 to the housing bubble of 2007, herding behavior has been
used to explain extreme market movements \citep{mackay:madness,schiller:bubble}. Most
of the literature discussed below focuses on models where the herd almost always leads
itself to an extreme outcome, where goods are blockbusters or flops.  Typically, agents
in these models have private information or preferences that are easily
drowned out by observing the behavior of others (and in some cases they have no private
information at all).  Conversely, the model here shows that when agents have
an evaluation strategy that is a mix of both private preferences and public actions
or information, then outcome distributions look much like that of day-to-day equity
returns: they may have kurtosis and skew that are arbitrarily large, but they remain
unimodal.  As the individual utility function is adjusted so that
private information is of little value, the model outcomes replicate the blockbusters,
flops, and market bifurcations in the literature.

Section \ref{litsec} will give a quick overview of the mostly empirical literature that
has demonstrated that equity returns are fat-tailed, and that equity traders (and those
who advise equity traders) demonstrate emulative behavior. 
\citet[p 20, emphasis in original.]{epstein:gas} wrote ``Perhaps one day people will interpret the question `Can you
explain it?' as asking `Can you {\em grow} it?'{}'' 
Section \ref{modelsec} will demonstrate that once we take emulative behavior as given,
it is easy to grow fat-tailed outcomes. Section \ref{endsec} concludes, pointing out 
that, because situations where outcomes are fat-tailed but not entirely off the charts are common,
we may be able to use emulative preferences to explain more than they have been used for
in the past.

\section{Literature}\label{litsec}

This section gives an overview of two threads of the economics
literature that do not quite meet. The first is an overview of the
existing literature on the distribution of equity returns; the second is
a survey of the situations posited in the finance literature where individuals
gain utility from emulating others.

\subsection{Fitting non-Normal distributions}
The second central moment, also known as the variance, is defined as:

$$\mu_2 = \sigma^2 = \int_{-\infty}^{\infty} (x-\mu)^2 f(x) dx,$$
where $x\in \Re$ is a random variable, 
$f(x)$ is the probability distribution function (PDF) on $x$,
and $\mu$ is the mean of $x$ $\left(\int_{-\infty}^{\infty} x f(x) dx\right)$.

One could similarly define the third and fourth central moments:

\begin{eqnarray*}
\mu_3 &=& \int_{-\infty}^{\infty} (x-\mu)^3 f(x) dx, \hbox{ and}\\
\mu_4 &=& \int_{-\infty}^{\infty} (x-\mu)^4 f(x) dx.
\end{eqnarray*}

Depending on the author, the {\em skew} is sometimes the third central moment,
$\skew \equiv \mu_3$, and sometimes $\skew \equiv \mu_3 / \sigma^{3}$. The {\em kurtosis} may be
$\kurt\equiv \mu_4$, $\kurt \equiv \mu_4/\sigma^4$, or $\kurt \equiv \mu_4/\sigma^4-3$.
In this paper, I will use 

\begin{eqnarray*}
\skew &\equiv& \mu_3 \hbox{, and}\\
\kurt &\equiv& \mu_4/\sigma^4.
\end{eqnarray*}

I will refer to $\kurt$ as {\em normalized kurtosis} to remind the reader that it is
divided by variance squared.

The more elaborate normalizations make it easy to compare these moments to a Normal
distribution, because for a Normal distribution with mean $\mu$ and standard deviation
$\sigma$,  $\mu_4/\sigma^4 = 3$.
A Normal distribution is symmetric and therefore has zero skew (whether normalized or not).
One can use these facts to
check empirical distributions for deviations from the Normal.

\citet{fama:kurt} ran such a test on equity returns\comment{ (price this period - price
last period)}, and found that they
were leptokurtic, meaning that $\mu_4 \gg 3 \sigma^4$, and were skewed.
However, he is not the first to notice these
features---\citet[footnote 3]{mandelbrot:certain} traces awareness of
the non-Normality of return distributions as far back as 1915.
Many of the papers cited in the following few paragraphs reproduce the results using their own
data sets.
\cite{bakshi:skew} gathered data on several
index and equity returns, and (with few exceptions) found a skew toward zero (i.e.,
negative skew, meaning that extreme downward events are more likely than extreme
upward events).

Most of the explanations for the deviation from the Normal have focused
on finding a closed-form PDF that better fits the
data. \citet{mandelbrot:certain} showed that a
stable Paretian (aka symmetric-stable) distribution fit better than the
Normal. \citet{blattberg:gonedes} showed that a renormalized Student's
$t$ distribution fit better than a symmetric-stable distribution.
\citet{kon:mixture} found that a mixture of Normal distributions fit better
than a Student's $t$. The mixture model produces an output distribution
by summing a first Normal distribution, ${\cal N}(\mu_1$, $\sigma_1)$,
with an independent second Normal distribution, ${\cal N}(\mu_2$, $\sigma_2)$.
Depending on the values of the five input parameters (two means, two
standard deviations, and a mixing parameter), the
distribution produced by summing the two can take on a wide range of mean,
standard deviation, skew, and kurtosis.

The mixture model raises a few critiques. Kon found that the sum of two distributions satisfactorily matches
only about half of the equity return distributions he tests. Others require as many
as four input distributions---and thus eleven input parameters---to explain the four
moments of the distribution to be matched.  \citet[pp 1095--96]{barb:clt} tested a
set of four broad equity indices (MSCI's USA, Europe, UK, and Japan indices) against a
comparable model claiming Normality with variances changing over time, and rejected the model for
all four indices. 

As with all of the distribution models, the use of a sum of several distributions
raises the question of how the given distributions go beyond being a good fit to being
a valid explanation of market behavior. After all, one could fit a Fourier sequence
to a data series to arbitrary precision, but it is not necessarily an explanation
of market behavior. This brings us to the second thread of the literature, covering
the micro-level behavior of market actors.

\subsection{Emulation} The literature provides many rational motivations for emulating 
others, variously termed herding, information cascades, network effects, peer effects,
spill\-overs---not to mention simple questions of fashion. 
This section provides a sample of some of the theoretical
results for such models, and a discussion of herding in the finance
context.
None of these models were written with the stated intention of describing an
observed leptokurtic distribution, but this section will calculate
the kurtosis of the output distributions implied by some of these
models to see how they fare.

\paragraph{The restaurant problem}		\label{rest}
Among the most common of the models where agents emulate others are
the {\em herding} or {\em information cascade} models, e.g. \cite{baner:herd}
or \cite{fads}. In these models, agents use the prior choices of other
agents as information when making decisions. 

A sequence of agents chooses to eat at restaurant $A$ or $B$. The first will use its
private information to choose. The second will use its private information, plus the
information revealed by the observable choice made by the first agent. The third agent
will add to its private information the information provided by observing where the first
two entrants are eating. Thus, if the first two agents are eating at restaurant $A$, the
third may ignore a preference for restaurant $B$ and eat at $A$. Once the preponderance
of prior choices leans toward restaurant $A$, we can expect that all future arrivals
will choose it as well. The next day, both restaurants start off empty again, and early
arrivals in the sequence might have private information that restaurant $B$ is better,
so subsequent arrivals would all go to restaurant $B$.

Network externalities are a property of goods where consumption by others increases the
utility of the good, such as a social networking web site whose utility depends on how
many others are also subscribed, computer equipment that needs to interoperate with
others' equipment, or coordination problems like the choice of whether to drive on the
right or left side of the road. The typical analysis (e.g., that of \cite{choi:herd})
matches that of the restaurant problem.

Both the information and the direct utility stories can be shown to produce a
bifurcated distribution of results with probability one: over many days, restaurant $A$
will show either about 0\% attendance or about 100\% attendance every day.  Many goods
show such a blockbuster/flop dichotomy, such as movies \citep{devaney:walls}.

But for our purposes, a sharply bimodal distribution is not desirable. First, one
would be hard-pressed to find an equity whose returns are truly bimodal.  
More importantly, such a bifurcated outcome distribution is typically {\em platykurtic},
the opposite of the leptokurtosis we seek. Consider an ideal bimodal distribution
with density $r \in (0, 1)$ at $a$ and density $1-r$ at $b$ (for any values of $a,
b\in \Re$, $a\neq b$). The distribution has
normalized kurtosis equal to $$\frac{1}{r-r^2} - 3.$$ For a symmetric
distribution, $r=0.5$, the normalized kurtosis is one, and it remains
less than three for any $r\in (.211, .789)$. Thus, a model that predicts
a bifurcated distribution can only show a large fourth moment if the distribution
is lopsided, which is not sustainable for equity returns.

\nocite{brock:durl}

\paragraph{Distribution models}
\cite{brock:durl} specify a model similar to the one presented here. In the first
round, a prior percentage of actors is given, and people act iff that percentage would
be large enough to give them a positive utility from acting.  In subsequent rounds,
individuals use the percentage of people who chose to act in the prior round to decide
whether to act or not.

The specific details of Brock and Durlauf's assumptions lead to two
possible outcomes. One is a bifurcation, much like the outcomes for
the restaurant problem models above. The other, due to the specific
form of the assumptions, is that the output distribution is the input
distribution transformed via the hyperbolic tangent. The
$\tanh$ transformation reduces the normalized kurtosis, and is therefore
inappropriate for deriving leptokurtic equity returns.

\citet{gss:crime} point out that the more people emulate others, the more
likely are extreme outcomes, which they measure via ``excess variance.''
They do this via a Binomial model: if being the victim of a crime is a draw from a Bernoulli
trial with probability $p$, then the mean of $n$ such trials is $np$, and the variance
is $np(1-p)$. Thus, given $n$ and the sample mean (or equivalently, $n$ and $p$) we can
solve for the expected variance, and if the observed variance is significantly greater,
then we can reject the hypothesis of independent Bernoulli trials. However,
this process says nothing about whether the observed victimization rates are Normally distributed or not:
excess variance is not excess kurtosis or skew.

\paragraph{Finance}

Within the theoretical finance literature, papers abound regarding herding behavior
(e.g., \cite{gross:stock, gross:re}, \cite{rad:reeq},
\cite{choi:herd}, \cite{minehart:herd}),
although they concern themselves not with explaining herding, but with
the information aggregation issues entailed by herding.
Many stories regarding the emulation of others apply to the
situation of the rational, self-interested manager of an asset portfolio:

\begin{itemize}
\item Pricing is partly based on the value of the underlying asset and
partly on what others are willing to pay for the asset. At the extreme, people 
will buy a stock which pays zero dividends only if they are confident
that there are other people who will also buy the stock; as more people
are willing to buy, the value of the stock to any individual rises.

\item It has long been a lament of the fund manager that if the herd
does badly but he breaks even, he sees little benefit; but if the herd
does well and he breaks even, then he gets fired.  Therefore, behaving
like others may explicitly appear in a risk-averse fund manager's utility
function. 

\item Since an undercapitalized company is likely to fail, the success
of a public offering may depend on how well-subscribed it is,
providing another justification for putting the behavior of others in
the fund manager's utility function.

\item If a large number of banks take simultaneous large losses, then they
may be bailed out; since a bail-out is unlikely if only one bank takes
a loss, this may also serve as an incentive for financiers to take
risks together.

\item Simply following the herd: ``[\dots] elements such as fashion and
sense of honour affected the banks' decision to take part in a syndicated
loan.  Banks are certainly not insensitive to prevailing trends, and
if it is `the in thing' to take part in syndicated loans[\dots], people
sometimes consent too readily.'' \cite[p 337]{jepma:macro}

\end{itemize}

The model of this paper is a reduced form model which simply
assumes that a financier's expected utility from an action is
increasing with the percentage of other people acting. I make no effort
to explain which of the above motivations are present at any
time, but assert that given these effects, the model below is applicable.

Empirical studies of analyst recommendations find that 
they do indeed herd. For example,
\cite{graham:analysts} finds evidence of herding among investment
newsletter recommendations, and finds that the {\sl more} reputable ones are
more likely to herd.  Meanwhile, \cite{hong:analysts} finds evidence
of herding among investment analysts, and finds that inexperienced
analysts are ``more likely to be terminated for bold forecasts that
deviate from consensus,'' and therefore {\sl less} reputable analysts are
more likely to herd.  \cite{welch:analysts} finds that an analyst
recommendation has a strong impact on the next two recommendations for
the same security by other analysts, and that this effect is uncorrelated
with whether the recommendations prove to be correct or not. Although
these papers disagree in the details, they all find 
empirical evidence that analysts are inclined to behave like other
analysts (and therefore the people who listen to analysts are likely to also 
behave alike), so the model below is apropos.

\section{The model}\label{modelsec}
One run of the model below finds an output equilibrium demand given an input
distribution of individual preferences. Repeating a single run 
thousands of times gives a distribution of equilibrium outcomes, which will
have large kurtosis and skew under certain conditions.

One run of the model consists of a plurality of agents (the simulations below use 10,000),
each privately deciding whether to purchase a good.
Each has an individual taste for consuming, $t\in \Re$, where $t\sim
{\cal N}(\epsilon,1)$ and $\epsilon$ is a small non-negative offset, fixed at
zero or 0.05 in the simulations to follow. 

Let the proportion of the population consuming be $k \in [0,1]$, and 
let the desire to emulate others be represented by a coefficient $\alpha\in
[0,\infty)$. Then the utility from consuming is 
\begin{equation}\label{uc}
U_c    = t + \alpha k.
\end{equation}
The utility from not consuming is 
\begin{equation}\label{unc}
U_{nc}    =  \alpha (1-k).
\end{equation}
That is, agents who do not consume get utility from emulating the $1-k$
agents who also do not consume, but have a taste for non-consumption
normalized to zero. One can show that this normalization is without loss
of generality. Agents consume iff $U_c > U_{nc}$.

A Bayesian Nash equilibrium is a set of acting agents, comprising the proportion $k_a$
percent of the population, where all acting agents have $U_c > U_{nc}$
given $k_a$ percent acting, and all agents outside the acting set have
$U_c \leq U_{nc}$ given $k_a$ percent acting.

\nocite{brock:durl}
It can be shown that, given the assumptions here, the game has a
cutoff-type equilibrium, where there is a cutoff value $T$ such that 
every agent with private tastes greater than $T$ acts and every agent
with $t \leq T$ does not act. An agent with private taste $t$ equal to
the cutoff $T$ will have $U_c = U_{nc}$.\comment{ One can show that there is exactly one 
equilibrium to every game (though its value will depend upon the individual values of $t$).\footnote{\citet{brock:durl}
assume a cutoff distribution for a situation such as the one here, 
but the existence of a unique cutoff can be derived from the
assumptions either in their paper or here.}}
\comment{
As a digression, the additional assumptions in Brock and Durlauf's paper
lead to a $\tanh$-form transformation of the base distribution to an
outcome distribution. The transformation {\em reduces} kurtosis, meaning
that their specific setup is inappropriate for the purpose at hand.}

One could embed this model of the distribution of demand into a larger model, such as
a simple supply-demand model where supply remains fixed and demand shifts as per the
model here, and prices thus vary with demand. To maintain focus on the core concept,
this paper will cover only the core model describing the distribution of $\hat k$.

\comment{\paragraph{Prices}
The key output variable from the model below is the percent acting. The next step would
be to use that percent acting to set the demand curve in a textbook pair of supply/demand
curves.  The supply curve would remain fixed, while the demand curve would translate
along the axis of units demanded as the percent buying the good rises or falls. This would
produce prices that rise and fall with demand, as expected. With linear supply/demand,
the translation induces a rescaling from percent acting to prices. Thus, from a series
of leptokurtic demands, a series of leptokurtic prices can easily follow. To keep the
model simple, the remainder of the paper will focus on generating leptokurtic demand,
without explicitly discussing the subsequent transformation from demand to prices.}

\subsection{Implementation}
Recall the restaurant problem, where we measured the turnout to restaurant $A$ every day
for a few weeks or months. Each day gave us another draw of diners from the population,
and it was the aggregate of turnouts over several days that added up to the bimodal
distribution.  Similarly, the literature on equities did not claim 
that if we surveyed willingness to pay by all members of the market at some instant
in time, the distribution would be leptokurtic; rather, the claim is that every day there is a new
distribution of willingness to pay, which produces a single outcome for the day,
and tallying those outcomes over time generates a leptokurtic distribution.
This model draws a sample distribution (which one could think of as today's
market, and which will be close to a Normal distribution), finds the equilibrium value
$\hat k$, and then repeats until there are enough samples of $\hat k$ that we can estimate
the moments of $\hat k$'s distribution.

We must first solve for the equilibrium percent acting for a single run.
Briefly switching from the equilibrium percent acting $k$ to the equilibrium cutoff taste
$T$, one can find the equilibrium for a single distribution by finding the value of
$T$ such that an agent with that value is indifferent between action and inaction,
given that the cutoff is at that value (that is, $U_c$ in Equation \ref{uc} equals
$U_{nc}$ in Equation \ref{unc}). Write the proportion not acting given cutoff $T$
as $\hbox{CDF}(T)$ (i.e. the cumulative distribution function of the empirical distribution of
tastes up to the cutoff $T$); then any value of $T$ that satisfies

\begin{equation} \label{popeq}
T = \alpha  (1 - 2\hbox{CDF}(T))
\end{equation}

is an equilibrium.

There are typically no closed-form solutions for $T$, so the work will require a numeric search.

I use an agent-based simulation to organize the draws. For each step, the simulation
draws 10,000 agents from the fixed distribution, then the simulation algorithm solves
for equilibrium via tat\^o{}nnement, as detailed below.  The equilibrium reached via
market simulation is a Bayesian Nash equilibrium as in Equation \ref{popeq}. There are
other search strategies for finding the equilibrium given the draws of $t$, but the
agent-based model has the advantages of always finding the equilibrium and providing
a realistic story of what happens in the market.

Repeating the process for thousands of draws from the fixed distribution, starting
each simulation with a new set of random draws of tastes $t$ from the same distribution,
will produce a distribution of the statistics $\hat T$ and $\hat k$,
including multiple modes when there are multiple equilibria.

The algorithm for a single run of the simulation is displayed in Figure
\ref{al}. In each step, agents consume or do not based on the value of
$k$ from the last step, and the process repeats until the value of $k$
no longer changes. The output of the process is the equilibrium value of
$\hat T$ and the equilibrium percent acting $\hat k$.

With a sufficiently large number of runs (in the simulations here, 20,000), it is
possible to calculate the moments $\skew(\hat k)$ and $\kurt(\hat k)$.

\begin{figure}
\vskip 0.5cm
--Fix $N$ and $\epsilon$.  \\
--Generate a new population of agents:  \\
\tabb--Set the initial value of $k=\frac{1}{2}$.\\
\tabb    --For each agent:  \\
\tabb\tabb--Draw a taste $t$ from a ${\cal N}(\epsilon,1)$
distribution. \\
--While $k$ this period is not equal to $k$ last period:  \\
\tabb    --For each agent:  \\
\tabb\tabb        --Consume iff $U_{k} \geq U_{nk}$.    \\
\tabb   --Recalculate $k$.   \\
--Record the equilibrium percent acting $k$.\\
\caption{The algorithm for finding the equilibrium level of consumption for one run.}
\label{al}
\end{figure}

The code itself is a short script written in C using the open source Apophenia library
\citep{klemens:modeling}, and is available upon request.

\begin{figure}

\begin{center}
\vspace{-2cm}
\vskip -2cm
\scalebox{1}{
  \input slices_c.tex
}

\vspace{-1cm}
\vskip -1cm

\scalebox{1.6}{\bbox{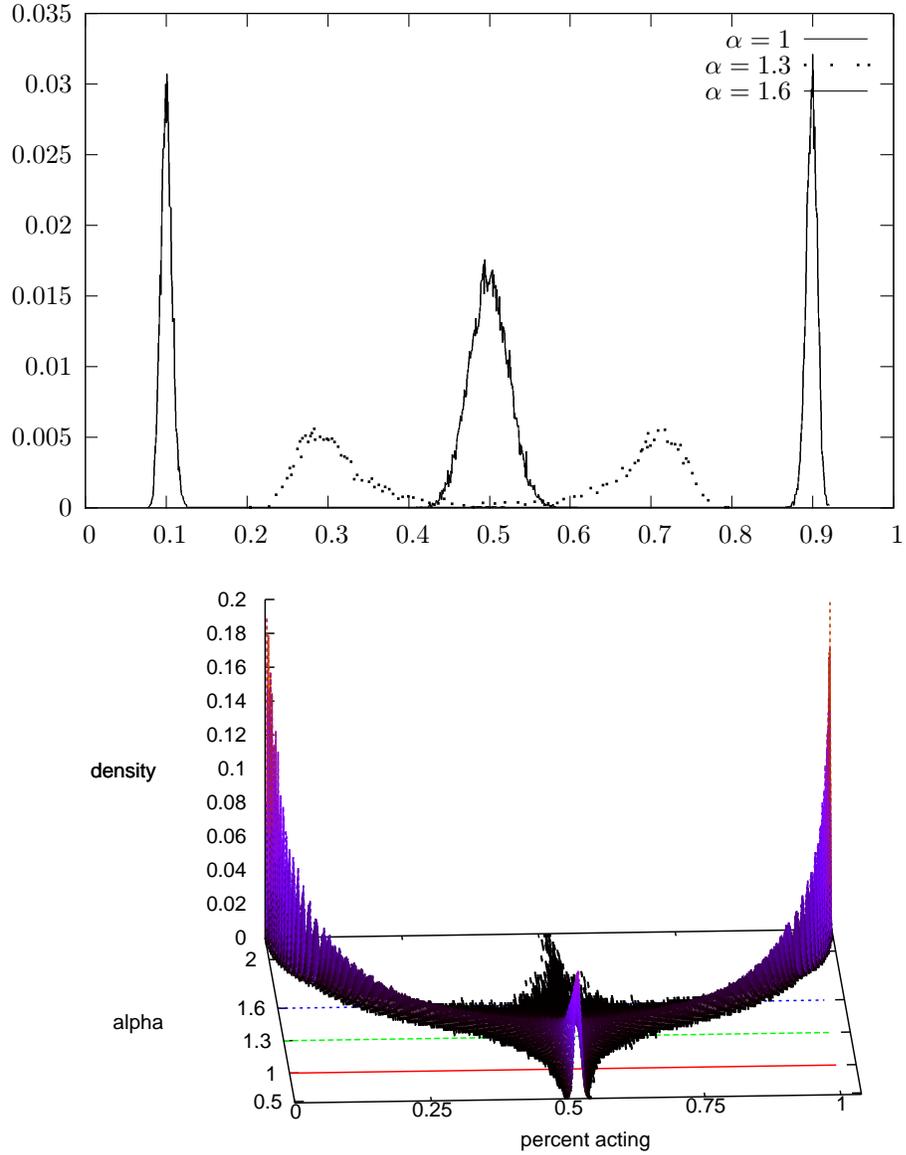}}
\end{center}
\vspace{-1cm}
\vskip -1cm

\caption{Above, three distributions of the equilibrium percent acting $k$,
for 20,000 runs with $\alpha$ equal to 1 (unimodal), 1.3 (bimodal with modes near 0.3 and
        0.7), and 1.6 (bimodal with modes near 0.1 and 0.9). Below, a
full sequence of such distributions, for $\alpha=0.5$ in front
up to $\alpha=2$ at the back. Vertical axis is the percent of runs (out
of 20,000 per $\alpha$) whose equilibrium is in the
given histogram bin. The three slices in the 2-D plot are indicated by a
line on the floor of the 3-D plot. }
\label{zero}
\end{figure}

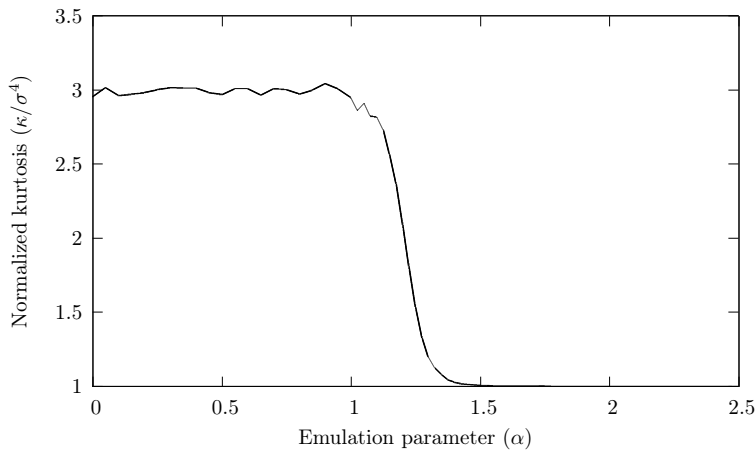
\begin{figure}
\begin{center}
\scalebox{0.80}{
\setlength{\unitlength}{0.240900pt}
\ifx\plotpoint\undefined\newsavebox{\plotpoint}\fi
\sbox{\plotpoint}{\rule[-0.200pt]{0.400pt}{0.400pt}}%
\begin{picture}(1500,900)(0,0)
\sbox{\plotpoint}{\rule[-0.200pt]{0.400pt}{0.400pt}}%
\put(171.0,131.0){\rule[-0.200pt]{4.818pt}{0.400pt}}
\put(151,131){\makebox(0,0)[r]{ 1}}
\put(1419.0,131.0){\rule[-0.200pt]{4.818pt}{0.400pt}}
\put(171.0,277.0){\rule[-0.200pt]{4.818pt}{0.400pt}}
\put(151,277){\makebox(0,0)[r]{ 1.5}}
\put(1419.0,277.0){\rule[-0.200pt]{4.818pt}{0.400pt}}
\put(171.0,422.0){\rule[-0.200pt]{4.818pt}{0.400pt}}
\put(151,422){\makebox(0,0)[r]{ 2}}
\put(1419.0,422.0){\rule[-0.200pt]{4.818pt}{0.400pt}}
\put(171.0,568.0){\rule[-0.200pt]{4.818pt}{0.400pt}}
\put(151,568){\makebox(0,0)[r]{ 2.5}}
\put(1419.0,568.0){\rule[-0.200pt]{4.818pt}{0.400pt}}
\put(171.0,713.0){\rule[-0.200pt]{4.818pt}{0.400pt}}
\put(151,713){\makebox(0,0)[r]{ 3}}
\put(1419.0,713.0){\rule[-0.200pt]{4.818pt}{0.400pt}}
\put(171.0,859.0){\rule[-0.200pt]{4.818pt}{0.400pt}}
\put(151,859){\makebox(0,0)[r]{ 3.5}}
\put(1419.0,859.0){\rule[-0.200pt]{4.818pt}{0.400pt}}
\put(171.0,131.0){\rule[-0.200pt]{0.400pt}{4.818pt}}
\put(171,90){\makebox(0,0){ 0}}
\put(171.0,839.0){\rule[-0.200pt]{0.400pt}{4.818pt}}
\put(425.0,131.0){\rule[-0.200pt]{0.400pt}{4.818pt}}
\put(425,90){\makebox(0,0){ 0.5}}
\put(425.0,839.0){\rule[-0.200pt]{0.400pt}{4.818pt}}
\put(678.0,131.0){\rule[-0.200pt]{0.400pt}{4.818pt}}
\put(678,90){\makebox(0,0){ 1}}
\put(678.0,839.0){\rule[-0.200pt]{0.400pt}{4.818pt}}
\put(932.0,131.0){\rule[-0.200pt]{0.400pt}{4.818pt}}
\put(932,90){\makebox(0,0){ 1.5}}
\put(932.0,839.0){\rule[-0.200pt]{0.400pt}{4.818pt}}
\put(1185.0,131.0){\rule[-0.200pt]{0.400pt}{4.818pt}}
\put(1185,90){\makebox(0,0){ 2}}
\put(1185.0,839.0){\rule[-0.200pt]{0.400pt}{4.818pt}}
\put(1439.0,131.0){\rule[-0.200pt]{0.400pt}{4.818pt}}
\put(1439,90){\makebox(0,0){ 2.5}}
\put(1439.0,839.0){\rule[-0.200pt]{0.400pt}{4.818pt}}
\put(171.0,131.0){\rule[-0.200pt]{0.400pt}{175.375pt}}
\put(171.0,131.0){\rule[-0.200pt]{305.461pt}{0.400pt}}
\put(1439.0,131.0){\rule[-0.200pt]{0.400pt}{175.375pt}}
\put(171.0,859.0){\rule[-0.200pt]{305.461pt}{0.400pt}}
\put(30,495){\makebox(0,0){\rotatebox{90}{Normalized kurtosis ($\kappa/\sigma^4$)}}}
\put(805,29){\makebox(0,0){Emulation parameter ($\alpha$)}}
\put(171,700){\usebox{\plotpoint}}
\multiput(171.00,700.58)(0.696,0.495){33}{\rule{0.656pt}{0.119pt}}
\multiput(171.00,699.17)(23.639,18.000){2}{\rule{0.328pt}{0.400pt}}
\multiput(196.00,716.92)(0.817,-0.494){29}{\rule{0.750pt}{0.119pt}}
\multiput(196.00,717.17)(24.443,-16.000){2}{\rule{0.375pt}{0.400pt}}
\multiput(222.00,702.61)(5.374,0.447){3}{\rule{3.433pt}{0.108pt}}
\multiput(222.00,701.17)(17.874,3.000){2}{\rule{1.717pt}{0.400pt}}
\multiput(247.00,705.61)(5.374,0.447){3}{\rule{3.433pt}{0.108pt}}
\multiput(247.00,704.17)(17.874,3.000){2}{\rule{1.717pt}{0.400pt}}
\multiput(272.00,708.59)(2.299,0.482){9}{\rule{1.833pt}{0.116pt}}
\multiput(272.00,707.17)(22.195,6.000){2}{\rule{0.917pt}{0.400pt}}
\multiput(298.00,714.60)(3.552,0.468){5}{\rule{2.600pt}{0.113pt}}
\multiput(298.00,713.17)(19.604,4.000){2}{\rule{1.300pt}{0.400pt}}
\put(323,716.67){\rule{6.263pt}{0.400pt}}
\multiput(323.00,717.17)(13.000,-1.000){2}{\rule{3.132pt}{0.400pt}}
\multiput(374.00,715.93)(1.427,-0.489){15}{\rule{1.211pt}{0.118pt}}
\multiput(374.00,716.17)(22.486,-9.000){2}{\rule{0.606pt}{0.400pt}}
\multiput(399.00,706.94)(3.698,-0.468){5}{\rule{2.700pt}{0.113pt}}
\multiput(399.00,707.17)(20.396,-4.000){2}{\rule{1.350pt}{0.400pt}}
\multiput(425.00,704.58)(1.056,0.492){21}{\rule{0.933pt}{0.119pt}}
\multiput(425.00,703.17)(23.063,12.000){2}{\rule{0.467pt}{0.400pt}}
\put(349.0,717.0){\rule[-0.200pt]{6.022pt}{0.400pt}}
\multiput(475.00,714.92)(1.012,-0.493){23}{\rule{0.900pt}{0.119pt}}
\multiput(475.00,715.17)(24.132,-13.000){2}{\rule{0.450pt}{0.400pt}}
\multiput(501.00,703.58)(0.972,0.493){23}{\rule{0.869pt}{0.119pt}}
\multiput(501.00,702.17)(23.196,13.000){2}{\rule{0.435pt}{0.400pt}}
\put(526,714.17){\rule{5.100pt}{0.400pt}}
\multiput(526.00,715.17)(14.415,-2.000){2}{\rule{2.550pt}{0.400pt}}
\multiput(551.00,712.93)(1.485,-0.489){15}{\rule{1.256pt}{0.118pt}}
\multiput(551.00,713.17)(23.394,-9.000){2}{\rule{0.628pt}{0.400pt}}
\multiput(577.00,705.59)(1.616,0.488){13}{\rule{1.350pt}{0.117pt}}
\multiput(577.00,704.17)(22.198,8.000){2}{\rule{0.675pt}{0.400pt}}
\multiput(602.00,713.58)(0.972,0.493){23}{\rule{0.869pt}{0.119pt}}
\multiput(602.00,712.17)(23.196,13.000){2}{\rule{0.435pt}{0.400pt}}
\multiput(627.00,724.92)(1.203,-0.492){19}{\rule{1.045pt}{0.118pt}}
\multiput(627.00,725.17)(23.830,-11.000){2}{\rule{0.523pt}{0.400pt}}
\multiput(653.00,713.92)(0.738,-0.495){31}{\rule{0.688pt}{0.119pt}}
\multiput(653.00,714.17)(23.572,-17.000){2}{\rule{0.344pt}{0.400pt}}
\multiput(678.58,694.39)(0.493,-0.972){23}{\rule{0.119pt}{0.869pt}}
\multiput(677.17,696.20)(13.000,-23.196){2}{\rule{0.400pt}{0.435pt}}
\multiput(691.58,673.00)(0.493,0.536){23}{\rule{0.119pt}{0.531pt}}
\multiput(690.17,673.00)(13.000,12.898){2}{\rule{0.400pt}{0.265pt}}
\multiput(704.58,683.13)(0.492,-1.056){21}{\rule{0.119pt}{0.933pt}}
\multiput(703.17,685.06)(12.000,-23.063){2}{\rule{0.400pt}{0.467pt}}
\put(716,660.17){\rule{2.700pt}{0.400pt}}
\multiput(716.00,661.17)(7.396,-2.000){2}{\rule{1.350pt}{0.400pt}}
\multiput(729.58,656.26)(0.493,-1.012){23}{\rule{0.119pt}{0.900pt}}
\multiput(728.17,658.13)(13.000,-24.132){2}{\rule{0.400pt}{0.450pt}}
\multiput(742.58,627.08)(0.492,-2.004){21}{\rule{0.119pt}{1.667pt}}
\multiput(741.17,630.54)(12.000,-43.541){2}{\rule{0.400pt}{0.833pt}}
\multiput(754.58,579.18)(0.493,-2.281){23}{\rule{0.119pt}{1.885pt}}
\multiput(753.17,583.09)(13.000,-54.088){2}{\rule{0.400pt}{0.942pt}}
\multiput(767.58,518.24)(0.493,-3.193){23}{\rule{0.119pt}{2.592pt}}
\multiput(766.17,523.62)(13.000,-75.620){2}{\rule{0.400pt}{1.296pt}}
\multiput(780.58,436.38)(0.492,-3.469){21}{\rule{0.119pt}{2.800pt}}
\multiput(779.17,442.19)(12.000,-75.188){2}{\rule{0.400pt}{1.400pt}}
\multiput(792.58,356.37)(0.493,-3.153){23}{\rule{0.119pt}{2.562pt}}
\multiput(791.17,361.68)(13.000,-74.683){2}{\rule{0.400pt}{1.281pt}}
\multiput(805.58,278.79)(0.493,-2.400){23}{\rule{0.119pt}{1.977pt}}
\multiput(804.17,282.90)(13.000,-56.897){2}{\rule{0.400pt}{0.988pt}}
\multiput(818.58,220.47)(0.492,-1.573){21}{\rule{0.119pt}{1.333pt}}
\multiput(817.17,223.23)(12.000,-34.233){2}{\rule{0.400pt}{0.667pt}}
\multiput(830.58,185.77)(0.493,-0.853){23}{\rule{0.119pt}{0.777pt}}
\multiput(829.17,187.39)(13.000,-20.387){2}{\rule{0.400pt}{0.388pt}}
\multiput(843.00,165.92)(0.497,-0.493){23}{\rule{0.500pt}{0.119pt}}
\multiput(843.00,166.17)(11.962,-13.000){2}{\rule{0.250pt}{0.400pt}}
\multiput(856.00,152.92)(0.600,-0.491){17}{\rule{0.580pt}{0.118pt}}
\multiput(856.00,153.17)(10.796,-10.000){2}{\rule{0.290pt}{0.400pt}}
\multiput(868.00,142.93)(1.378,-0.477){7}{\rule{1.140pt}{0.115pt}}
\multiput(868.00,143.17)(10.634,-5.000){2}{\rule{0.570pt}{0.400pt}}
\multiput(881.00,137.95)(2.695,-0.447){3}{\rule{1.833pt}{0.108pt}}
\multiput(881.00,138.17)(9.195,-3.000){2}{\rule{0.917pt}{0.400pt}}
\put(894,134.67){\rule{2.891pt}{0.400pt}}
\multiput(894.00,135.17)(6.000,-1.000){2}{\rule{1.445pt}{0.400pt}}
\put(906,133.67){\rule{3.132pt}{0.400pt}}
\multiput(906.00,134.17)(6.500,-1.000){2}{\rule{1.566pt}{0.400pt}}
\put(919,132.67){\rule{3.132pt}{0.400pt}}
\multiput(919.00,133.17)(6.500,-1.000){2}{\rule{1.566pt}{0.400pt}}
\put(450.0,716.0){\rule[-0.200pt]{6.022pt}{0.400pt}}
\put(944,131.67){\rule{3.132pt}{0.400pt}}
\multiput(944.00,132.17)(6.500,-1.000){2}{\rule{1.566pt}{0.400pt}}
\put(932.0,133.0){\rule[-0.200pt]{2.891pt}{0.400pt}}
\put(1046,130.67){\rule{6.023pt}{0.400pt}}
\multiput(1046.00,131.17)(12.500,-1.000){2}{\rule{3.011pt}{0.400pt}}
\put(957.0,132.0){\rule[-0.200pt]{21.440pt}{0.400pt}}
\put(1071.0,131.0){\rule[-0.200pt]{85.519pt}{0.400pt}}
\put(171.0,131.0){\rule[-0.200pt]{0.400pt}{175.375pt}}
\put(171.0,131.0){\rule[-0.200pt]{305.461pt}{0.400pt}}
\put(1439.0,131.0){\rule[-0.200pt]{0.400pt}{175.375pt}}
\put(171.0,859.0){\rule[-0.200pt]{305.461pt}{0.400pt}}
\end{picture}
}
\end{center}
\caption{The normalized kurtosis reveals the narrow range of transition
from Normal-type distribution ($\kappa/\sigma^4=3$) to
bimodal-type distribution
($\kappa/\sigma^4=1$).}
\label{kzero}
\end{figure}

\subsection{Results}
It is instructive to begin with the symmetric case, where $\epsilon=0$, so agents' private
tastes are drawn from a ${\cal N}(0, 1)$ distribution. 

Figure \ref{zero} shows a sequence of distributions of the equilibrium
percent acting $k$, from the distribution given
$\alpha=1$ up to the distribution for $\alpha=2.5$, with distributions
for three specific values of $\alpha$ highlighted. 
Small values of $\alpha$ (where utility is mostly private valuation)
result in a Normal output distribution of prices, while large values of
$\alpha$ (where utility is mostly public) give a
coordination-game style bifurcation. 

As $\alpha$ goes from the Normal range to the bifurcated range, there is a small range of $\alpha$ where
the transition occurs, and the distribution is neither fully bifurcated nor Normal.

At large $\alpha$, the value of $k$ between the sink that sends the simulation to
the lower equilibrium and the sink that sends the simulation to the higher equilibrium 
(near 0.5) is an unstable equilibrium; in theory it occurs with probability zero, but in a finite
simulation it occurs with small probability.\footnote{The figures are the aggregate of
20,000 runs of the simulation. If an equilibrium was reached even once, then it appears as a mark in the 
3-D plot. The 2-D plots have lower resolution, and unlikely events may blend with the axes.} 
Below, we will see that these distributions with a small middle mode behave like a
bifurcated distribution, so I will refer to them as such.

The small transition range is especially clear when we look at the normalized kurtosis
of each $\alpha$'s distribution, which is not at all a uniform shift. As in Figure
\ref{kzero}, the normalized kurtosis is consistently three for small values of $\alpha$
(as for a Normal distribution), is consistently one for large values of $\alpha$ (as
for a symmetric bimodal distribution),  and has a quick period of transition
between $\alpha \approx 1$ and $\alpha \approx 1.4$.\footnote{The units
on $\alpha$ are utils per percent acting, so exact values of $\alpha$ are basically
meaningless. Rescaling $t$ (by changing its variance) would produce entirely different
values of $\alpha$, but the qualitative effects described here would still hold.}

\begin{figure}
\begin{center}
\vspace{-2cm}
\vskip -2cm
\scalebox{1}{
\input slices_oc.tex
}
\vspace{-1cm}
\vskip -1cm

\scalebox{1.6}{\bbox{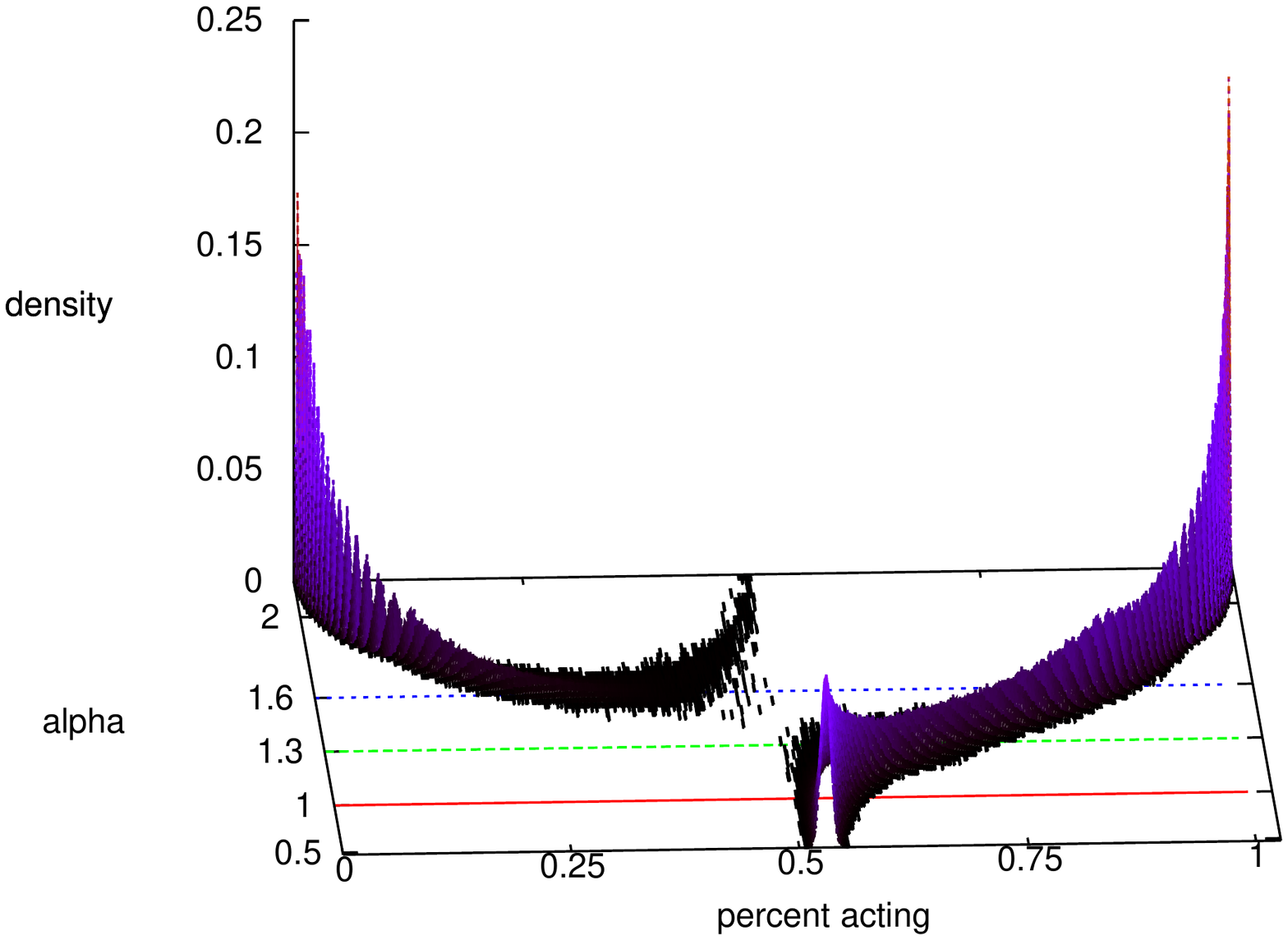}}
\end{center}
\caption{Two views of the $\alpha$-to-cutoff-frequency relation. 
    PDFs of cutoffs for three given levels of $\alpha$ 
    (1=unimodal near center, 1.3=unimodal to right, 1.6=bimodal) are displayed in 2-D form at top.
    At bottom is a series of PDFs for a range of values of $\alpha$ from
    $\alpha=0.5$ at front to $\alpha=2$ at rear.}
\label{plot}
\end{figure}

\begin{figure}
\begin{center}
\scalebox{1.00}{
\setlength{\unitlength}{0.240900pt}
\ifx\plotpoint\undefined\newsavebox{\plotpoint}\fi
\sbox{\plotpoint}{\rule[-0.200pt]{0.400pt}{0.400pt}}%
\begin{picture}(1500,900)(0,0)
\sbox{\plotpoint}{\rule[-0.200pt]{0.400pt}{0.400pt}}%
\put(171.0,131.0){\rule[-0.200pt]{4.818pt}{0.400pt}}
\put(151,131){\makebox(0,0)[r]{ 0}}
\put(1419.0,131.0){\rule[-0.200pt]{4.818pt}{0.400pt}}
\put(171.0,277.0){\rule[-0.200pt]{4.818pt}{0.400pt}}
\put(151,277){\makebox(0,0)[r]{ 50}}
\put(1419.0,277.0){\rule[-0.200pt]{4.818pt}{0.400pt}}
\put(171.0,422.0){\rule[-0.200pt]{4.818pt}{0.400pt}}
\put(151,422){\makebox(0,0)[r]{ 100}}
\put(1419.0,422.0){\rule[-0.200pt]{4.818pt}{0.400pt}}
\put(171.0,568.0){\rule[-0.200pt]{4.818pt}{0.400pt}}
\put(151,568){\makebox(0,0)[r]{ 150}}
\put(1419.0,568.0){\rule[-0.200pt]{4.818pt}{0.400pt}}
\put(171.0,713.0){\rule[-0.200pt]{4.818pt}{0.400pt}}
\put(151,713){\makebox(0,0)[r]{ 200}}
\put(1419.0,713.0){\rule[-0.200pt]{4.818pt}{0.400pt}}
\put(171.0,859.0){\rule[-0.200pt]{4.818pt}{0.400pt}}
\put(151,859){\makebox(0,0)[r]{ 250}}
\put(1419.0,859.0){\rule[-0.200pt]{4.818pt}{0.400pt}}
\put(171.0,131.0){\rule[-0.200pt]{0.400pt}{4.818pt}}
\put(171,90){\makebox(0,0){ 0}}
\put(171.0,839.0){\rule[-0.200pt]{0.400pt}{4.818pt}}
\put(425.0,131.0){\rule[-0.200pt]{0.400pt}{4.818pt}}
\put(425,90){\makebox(0,0){ 0.5}}
\put(425.0,839.0){\rule[-0.200pt]{0.400pt}{4.818pt}}
\put(678.0,131.0){\rule[-0.200pt]{0.400pt}{4.818pt}}
\put(678,90){\makebox(0,0){ 1}}
\put(678.0,839.0){\rule[-0.200pt]{0.400pt}{4.818pt}}
\put(932.0,131.0){\rule[-0.200pt]{0.400pt}{4.818pt}}
\put(932,90){\makebox(0,0){ 1.5}}
\put(932.0,839.0){\rule[-0.200pt]{0.400pt}{4.818pt}}
\put(1185.0,131.0){\rule[-0.200pt]{0.400pt}{4.818pt}}
\put(1185,90){\makebox(0,0){ 2}}
\put(1185.0,839.0){\rule[-0.200pt]{0.400pt}{4.818pt}}
\put(1439.0,131.0){\rule[-0.200pt]{0.400pt}{4.818pt}}
\put(1439,90){\makebox(0,0){ 2.5}}
\put(1439.0,839.0){\rule[-0.200pt]{0.400pt}{4.818pt}}
\put(171.0,131.0){\rule[-0.200pt]{0.400pt}{175.375pt}}
\put(171.0,131.0){\rule[-0.200pt]{305.461pt}{0.400pt}}
\put(1439.0,131.0){\rule[-0.200pt]{0.400pt}{175.375pt}}
\put(171.0,859.0){\rule[-0.200pt]{305.461pt}{0.400pt}}
\put(30,495){\makebox(0,0){\rotatebox{90}{Normalized kurtosis ($\kappa/\sigma^4$)}}}
\put(805,29){\makebox(0,0){Emulation parameter ($\alpha$)}}
\put(171,140){\usebox{\plotpoint}}
\put(729,139.67){\rule{3.132pt}{0.400pt}}
\multiput(729.00,139.17)(6.500,1.000){2}{\rule{1.566pt}{0.400pt}}
\put(742,140.67){\rule{2.891pt}{0.400pt}}
\multiput(742.00,140.17)(6.000,1.000){2}{\rule{1.445pt}{0.400pt}}
\put(171.0,140.0){\rule[-0.200pt]{134.422pt}{0.400pt}}
\put(767,142.17){\rule{2.700pt}{0.400pt}}
\multiput(767.00,141.17)(7.396,2.000){2}{\rule{1.350pt}{0.400pt}}
\multiput(780.00,144.61)(2.472,0.447){3}{\rule{1.700pt}{0.108pt}}
\multiput(780.00,143.17)(8.472,3.000){2}{\rule{0.850pt}{0.400pt}}
\put(792,145.17){\rule{2.700pt}{0.400pt}}
\multiput(792.00,146.17)(7.396,-2.000){2}{\rule{1.350pt}{0.400pt}}
\multiput(805.58,145.00)(0.493,3.272){23}{\rule{0.119pt}{2.654pt}}
\multiput(804.17,145.00)(13.000,77.492){2}{\rule{0.400pt}{1.327pt}}
\multiput(818.58,228.00)(0.492,10.750){21}{\rule{0.119pt}{8.433pt}}
\multiput(817.17,228.00)(12.000,232.496){2}{\rule{0.400pt}{4.217pt}}
\multiput(830.58,478.00)(0.493,11.162){23}{\rule{0.119pt}{8.777pt}}
\multiput(829.17,478.00)(13.000,263.783){2}{\rule{0.400pt}{4.388pt}}
\multiput(843.58,712.45)(0.493,-14.612){23}{\rule{0.119pt}{11.454pt}}
\multiput(842.17,736.23)(13.000,-345.227){2}{\rule{0.400pt}{5.727pt}}
\multiput(856.58,364.99)(0.492,-7.949){21}{\rule{0.119pt}{6.267pt}}
\multiput(855.17,377.99)(12.000,-171.993){2}{\rule{0.400pt}{3.133pt}}
\multiput(868.58,199.33)(0.493,-1.924){23}{\rule{0.119pt}{1.608pt}}
\multiput(867.17,202.66)(13.000,-45.663){2}{\rule{0.400pt}{0.804pt}}
\multiput(881.58,154.54)(0.493,-0.616){23}{\rule{0.119pt}{0.592pt}}
\multiput(880.17,155.77)(13.000,-14.771){2}{\rule{0.400pt}{0.296pt}}
\multiput(894.00,139.93)(1.267,-0.477){7}{\rule{1.060pt}{0.115pt}}
\multiput(894.00,140.17)(9.800,-5.000){2}{\rule{0.530pt}{0.400pt}}
\put(906,134.67){\rule{3.132pt}{0.400pt}}
\multiput(906.00,135.17)(6.500,-1.000){2}{\rule{1.566pt}{0.400pt}}
\put(754.0,142.0){\rule[-0.200pt]{3.132pt}{0.400pt}}
\put(932,133.67){\rule{2.891pt}{0.400pt}}
\multiput(932.00,134.17)(6.000,-1.000){2}{\rule{1.445pt}{0.400pt}}
\put(919.0,135.0){\rule[-0.200pt]{3.132pt}{0.400pt}}
\put(944.0,134.0){\rule[-0.200pt]{116.114pt}{0.400pt}}
\put(171.0,131.0){\rule[-0.200pt]{0.400pt}{175.375pt}}
\put(171.0,131.0){\rule[-0.200pt]{305.461pt}{0.400pt}}
\put(1439.0,131.0){\rule[-0.200pt]{0.400pt}{175.375pt}}
\put(171.0,859.0){\rule[-0.200pt]{305.461pt}{0.400pt}}
\end{picture}
}
\scalebox{1.00}{
\setlength{\unitlength}{0.240900pt}
\ifx\plotpoint\undefined\newsavebox{\plotpoint}\fi
\sbox{\plotpoint}{\rule[-0.200pt]{0.400pt}{0.400pt}}%
\begin{picture}(1500,900)(0,0)
\sbox{\plotpoint}{\rule[-0.200pt]{0.400pt}{0.400pt}}%
\put(151.0,131.0){\rule[-0.200pt]{4.818pt}{0.400pt}}
\put(131,131){\makebox(0,0)[r]{-12}}
\put(1419.0,131.0){\rule[-0.200pt]{4.818pt}{0.400pt}}
\put(151.0,235.0){\rule[-0.200pt]{4.818pt}{0.400pt}}
\put(131,235){\makebox(0,0)[r]{-10}}
\put(1419.0,235.0){\rule[-0.200pt]{4.818pt}{0.400pt}}
\put(151.0,339.0){\rule[-0.200pt]{4.818pt}{0.400pt}}
\put(131,339){\makebox(0,0)[r]{-8}}
\put(1419.0,339.0){\rule[-0.200pt]{4.818pt}{0.400pt}}
\put(151.0,443.0){\rule[-0.200pt]{4.818pt}{0.400pt}}
\put(131,443){\makebox(0,0)[r]{-6}}
\put(1419.0,443.0){\rule[-0.200pt]{4.818pt}{0.400pt}}
\put(151.0,547.0){\rule[-0.200pt]{4.818pt}{0.400pt}}
\put(131,547){\makebox(0,0)[r]{-4}}
\put(1419.0,547.0){\rule[-0.200pt]{4.818pt}{0.400pt}}
\put(151.0,651.0){\rule[-0.200pt]{4.818pt}{0.400pt}}
\put(131,651){\makebox(0,0)[r]{-2}}
\put(1419.0,651.0){\rule[-0.200pt]{4.818pt}{0.400pt}}
\put(151.0,755.0){\rule[-0.200pt]{4.818pt}{0.400pt}}
\put(131,755){\makebox(0,0)[r]{ 0}}
\put(1419.0,755.0){\rule[-0.200pt]{4.818pt}{0.400pt}}
\put(151.0,859.0){\rule[-0.200pt]{4.818pt}{0.400pt}}
\put(131,859){\makebox(0,0)[r]{ 2}}
\put(1419.0,859.0){\rule[-0.200pt]{4.818pt}{0.400pt}}
\put(151.0,131.0){\rule[-0.200pt]{0.400pt}{4.818pt}}
\put(151,90){\makebox(0,0){ 0}}
\put(151.0,839.0){\rule[-0.200pt]{0.400pt}{4.818pt}}
\put(409.0,131.0){\rule[-0.200pt]{0.400pt}{4.818pt}}
\put(409,90){\makebox(0,0){ 0.5}}
\put(409.0,839.0){\rule[-0.200pt]{0.400pt}{4.818pt}}
\put(666.0,131.0){\rule[-0.200pt]{0.400pt}{4.818pt}}
\put(666,90){\makebox(0,0){ 1}}
\put(666.0,839.0){\rule[-0.200pt]{0.400pt}{4.818pt}}
\put(924.0,131.0){\rule[-0.200pt]{0.400pt}{4.818pt}}
\put(924,90){\makebox(0,0){ 1.5}}
\put(924.0,839.0){\rule[-0.200pt]{0.400pt}{4.818pt}}
\put(1181.0,131.0){\rule[-0.200pt]{0.400pt}{4.818pt}}
\put(1181,90){\makebox(0,0){ 2}}
\put(1181.0,839.0){\rule[-0.200pt]{0.400pt}{4.818pt}}
\put(1439.0,131.0){\rule[-0.200pt]{0.400pt}{4.818pt}}
\put(1439,90){\makebox(0,0){ 2.5}}
\put(1439.0,839.0){\rule[-0.200pt]{0.400pt}{4.818pt}}
\put(151.0,131.0){\rule[-0.200pt]{0.400pt}{175.375pt}}
\put(151.0,131.0){\rule[-0.200pt]{310.279pt}{0.400pt}}
\put(1439.0,131.0){\rule[-0.200pt]{0.400pt}{175.375pt}}
\put(151.0,859.0){\rule[-0.200pt]{310.279pt}{0.400pt}}
\put(30,495){\makebox(0,0){\rotatebox{90}{Normalized skew ($\skew/\sigma^3$)}}}
\put(795,29){\makebox(0,0){Emulation parameter ($\alpha$)}}
\put(151,756){\usebox{\plotpoint}}
\put(177,754.67){\rule{6.263pt}{0.400pt}}
\multiput(177.00,755.17)(13.000,-1.000){2}{\rule{3.132pt}{0.400pt}}
\put(203,753.67){\rule{6.023pt}{0.400pt}}
\multiput(203.00,754.17)(12.500,-1.000){2}{\rule{3.011pt}{0.400pt}}
\put(228,753.67){\rule{6.263pt}{0.400pt}}
\multiput(228.00,753.17)(13.000,1.000){2}{\rule{3.132pt}{0.400pt}}
\put(254,753.67){\rule{6.263pt}{0.400pt}}
\multiput(254.00,754.17)(13.000,-1.000){2}{\rule{3.132pt}{0.400pt}}
\put(280,754.17){\rule{5.300pt}{0.400pt}}
\multiput(280.00,753.17)(15.000,2.000){2}{\rule{2.650pt}{0.400pt}}
\put(306,754.67){\rule{6.023pt}{0.400pt}}
\multiput(306.00,755.17)(12.500,-1.000){2}{\rule{3.011pt}{0.400pt}}
\put(151.0,756.0){\rule[-0.200pt]{6.263pt}{0.400pt}}
\put(357,754.67){\rule{6.263pt}{0.400pt}}
\multiput(357.00,754.17)(13.000,1.000){2}{\rule{3.132pt}{0.400pt}}
\put(383,754.17){\rule{5.300pt}{0.400pt}}
\multiput(383.00,755.17)(15.000,-2.000){2}{\rule{2.650pt}{0.400pt}}
\put(409,754.17){\rule{5.100pt}{0.400pt}}
\multiput(409.00,753.17)(14.415,2.000){2}{\rule{2.550pt}{0.400pt}}
\multiput(434.00,754.95)(5.597,-0.447){3}{\rule{3.567pt}{0.108pt}}
\multiput(434.00,755.17)(18.597,-3.000){2}{\rule{1.783pt}{0.400pt}}
\put(460,753.17){\rule{5.300pt}{0.400pt}}
\multiput(460.00,752.17)(15.000,2.000){2}{\rule{2.650pt}{0.400pt}}
\put(486,753.17){\rule{5.300pt}{0.400pt}}
\multiput(486.00,754.17)(15.000,-2.000){2}{\rule{2.650pt}{0.400pt}}
\put(512,751.67){\rule{6.023pt}{0.400pt}}
\multiput(512.00,752.17)(12.500,-1.000){2}{\rule{3.011pt}{0.400pt}}
\put(537,751.67){\rule{6.263pt}{0.400pt}}
\multiput(537.00,751.17)(13.000,1.000){2}{\rule{3.132pt}{0.400pt}}
\put(563,751.17){\rule{5.300pt}{0.400pt}}
\multiput(563.00,752.17)(15.000,-2.000){2}{\rule{2.650pt}{0.400pt}}
\put(589,750.67){\rule{6.263pt}{0.400pt}}
\multiput(589.00,750.17)(13.000,1.000){2}{\rule{3.132pt}{0.400pt}}
\put(615,751.67){\rule{6.023pt}{0.400pt}}
\multiput(615.00,751.17)(12.500,1.000){2}{\rule{3.011pt}{0.400pt}}
\multiput(640.00,751.95)(5.597,-0.447){3}{\rule{3.567pt}{0.108pt}}
\multiput(640.00,752.17)(18.597,-3.000){2}{\rule{1.783pt}{0.400pt}}
\multiput(666.00,748.93)(1.378,-0.477){7}{\rule{1.140pt}{0.115pt}}
\multiput(666.00,749.17)(10.634,-5.000){2}{\rule{0.570pt}{0.400pt}}
\multiput(679.00,743.95)(2.695,-0.447){3}{\rule{1.833pt}{0.108pt}}
\multiput(679.00,744.17)(9.195,-3.000){2}{\rule{0.917pt}{0.400pt}}
\put(692,740.17){\rule{2.700pt}{0.400pt}}
\multiput(692.00,741.17)(7.396,-2.000){2}{\rule{1.350pt}{0.400pt}}
\multiput(705.00,738.94)(1.797,-0.468){5}{\rule{1.400pt}{0.113pt}}
\multiput(705.00,739.17)(10.094,-4.000){2}{\rule{0.700pt}{0.400pt}}
\multiput(718.00,734.93)(1.378,-0.477){7}{\rule{1.140pt}{0.115pt}}
\multiput(718.00,735.17)(10.634,-5.000){2}{\rule{0.570pt}{0.400pt}}
\multiput(731.00,729.93)(0.874,-0.485){11}{\rule{0.786pt}{0.117pt}}
\multiput(731.00,730.17)(10.369,-7.000){2}{\rule{0.393pt}{0.400pt}}
\put(743,722.17){\rule{2.700pt}{0.400pt}}
\multiput(743.00,723.17)(7.396,-2.000){2}{\rule{1.350pt}{0.400pt}}
\multiput(756.00,720.93)(0.824,-0.488){13}{\rule{0.750pt}{0.117pt}}
\multiput(756.00,721.17)(11.443,-8.000){2}{\rule{0.375pt}{0.400pt}}
\put(769,712.17){\rule{2.700pt}{0.400pt}}
\multiput(769.00,713.17)(7.396,-2.000){2}{\rule{1.350pt}{0.400pt}}
\multiput(782.00,712.61)(2.695,0.447){3}{\rule{1.833pt}{0.108pt}}
\multiput(782.00,711.17)(9.195,3.000){2}{\rule{0.917pt}{0.400pt}}
\multiput(795.58,705.39)(0.493,-2.836){23}{\rule{0.119pt}{2.315pt}}
\multiput(794.17,710.19)(13.000,-67.194){2}{\rule{0.400pt}{1.158pt}}
\multiput(808.58,617.42)(0.493,-7.792){23}{\rule{0.119pt}{6.162pt}}
\multiput(807.17,630.21)(13.000,-184.211){2}{\rule{0.400pt}{3.081pt}}
\multiput(821.58,409.31)(0.493,-11.241){23}{\rule{0.119pt}{8.838pt}}
\multiput(820.17,427.66)(13.000,-265.655){2}{\rule{0.400pt}{4.419pt}}
\multiput(834.58,162.00)(0.493,4.937){23}{\rule{0.119pt}{3.946pt}}
\multiput(833.17,162.00)(13.000,116.810){2}{\rule{0.400pt}{1.973pt}}
\multiput(847.58,287.00)(0.492,9.113){21}{\rule{0.119pt}{7.167pt}}
\multiput(846.17,287.00)(12.000,197.125){2}{\rule{0.400pt}{3.583pt}}
\multiput(859.58,499.00)(0.493,4.343){23}{\rule{0.119pt}{3.485pt}}
\multiput(858.17,499.00)(13.000,102.768){2}{\rule{0.400pt}{1.742pt}}
\multiput(872.58,609.00)(0.493,2.479){23}{\rule{0.119pt}{2.038pt}}
\multiput(871.17,609.00)(13.000,58.769){2}{\rule{0.400pt}{1.019pt}}
\multiput(885.58,672.00)(0.493,1.408){23}{\rule{0.119pt}{1.208pt}}
\multiput(884.17,672.00)(13.000,33.493){2}{\rule{0.400pt}{0.604pt}}
\multiput(898.58,708.00)(0.493,0.616){23}{\rule{0.119pt}{0.592pt}}
\multiput(897.17,708.00)(13.000,14.771){2}{\rule{0.400pt}{0.296pt}}
\multiput(911.00,724.59)(1.123,0.482){9}{\rule{0.967pt}{0.116pt}}
\multiput(911.00,723.17)(10.994,6.000){2}{\rule{0.483pt}{0.400pt}}
\multiput(924.00,730.60)(1.797,0.468){5}{\rule{1.400pt}{0.113pt}}
\multiput(924.00,729.17)(10.094,4.000){2}{\rule{0.700pt}{0.400pt}}
\put(331.0,755.0){\rule[-0.200pt]{6.263pt}{0.400pt}}
\multiput(950.00,734.59)(1.267,0.477){7}{\rule{1.060pt}{0.115pt}}
\multiput(950.00,733.17)(9.800,5.000){2}{\rule{0.530pt}{0.400pt}}
\put(937.0,734.0){\rule[-0.200pt]{3.132pt}{0.400pt}}
\put(975,739.17){\rule{2.700pt}{0.400pt}}
\multiput(975.00,738.17)(7.396,2.000){2}{\rule{1.350pt}{0.400pt}}
\put(988,740.67){\rule{6.263pt}{0.400pt}}
\multiput(988.00,740.17)(13.000,1.000){2}{\rule{3.132pt}{0.400pt}}
\put(1014,742.17){\rule{5.300pt}{0.400pt}}
\multiput(1014.00,741.17)(15.000,2.000){2}{\rule{2.650pt}{0.400pt}}
\put(962.0,739.0){\rule[-0.200pt]{3.132pt}{0.400pt}}
\put(1065,744.17){\rule{5.300pt}{0.400pt}}
\multiput(1065.00,743.17)(15.000,2.000){2}{\rule{2.650pt}{0.400pt}}
\put(1091,745.67){\rule{6.263pt}{0.400pt}}
\multiput(1091.00,745.17)(13.000,1.000){2}{\rule{3.132pt}{0.400pt}}
\put(1040.0,744.0){\rule[-0.200pt]{6.022pt}{0.400pt}}
\put(1169,747.17){\rule{5.100pt}{0.400pt}}
\multiput(1169.00,746.17)(14.415,2.000){2}{\rule{2.550pt}{0.400pt}}
\put(1194,748.67){\rule{6.263pt}{0.400pt}}
\multiput(1194.00,748.17)(13.000,1.000){2}{\rule{3.132pt}{0.400pt}}
\put(1220,748.67){\rule{6.263pt}{0.400pt}}
\multiput(1220.00,749.17)(13.000,-1.000){2}{\rule{3.132pt}{0.400pt}}
\put(1246,749.17){\rule{5.300pt}{0.400pt}}
\multiput(1246.00,748.17)(15.000,2.000){2}{\rule{2.650pt}{0.400pt}}
\put(1272,749.17){\rule{5.100pt}{0.400pt}}
\multiput(1272.00,750.17)(14.415,-2.000){2}{\rule{2.550pt}{0.400pt}}
\put(1297,749.17){\rule{5.300pt}{0.400pt}}
\multiput(1297.00,748.17)(15.000,2.000){2}{\rule{2.650pt}{0.400pt}}
\put(1117.0,747.0){\rule[-0.200pt]{12.527pt}{0.400pt}}
\put(1349,749.67){\rule{6.263pt}{0.400pt}}
\multiput(1349.00,750.17)(13.000,-1.000){2}{\rule{3.132pt}{0.400pt}}
\put(1323.0,751.0){\rule[-0.200pt]{6.263pt}{0.400pt}}
\put(1375.0,750.0){\rule[-0.200pt]{12.286pt}{0.400pt}}
\put(151.0,131.0){\rule[-0.200pt]{0.400pt}{175.375pt}}
\put(151.0,131.0){\rule[-0.200pt]{310.279pt}{0.400pt}}
\put(1439.0,131.0){\rule[-0.200pt]{0.400pt}{175.375pt}}
\put(151.0,859.0){\rule[-0.200pt]{310.279pt}{0.400pt}}
\end{picture}
}

\end{center}
\caption{The relationship between $n$ (on the horizontal axis) 
and $\kurt/\sigma^4$ (on the vertical axis, top) or $\skew/\sigma^3$ (on the vertical axis, bottom).}
\label{ksplot}
\end{figure}
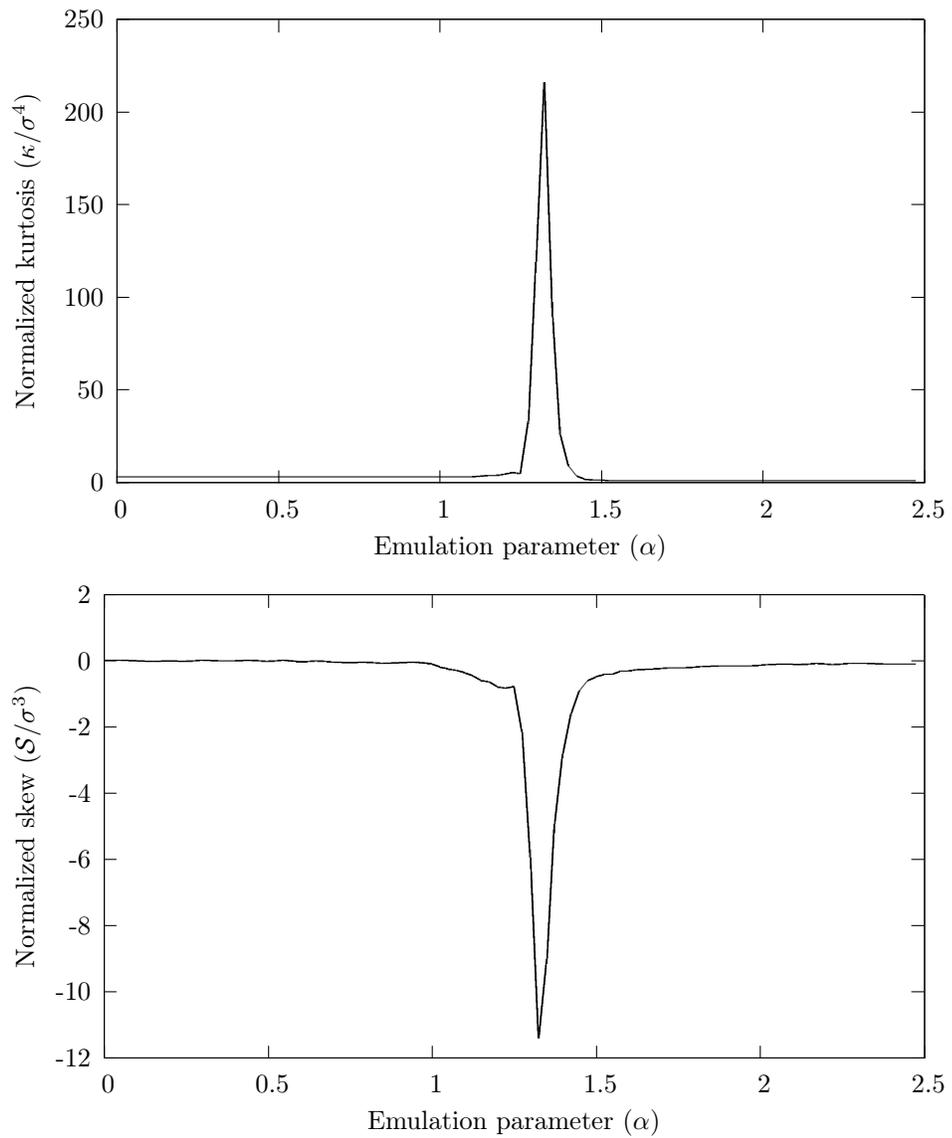

Figure \ref{plot} shows the sequence of distributions where $\epsilon=0.05$.
For $\alpha\approx 0$, the
distribution is roughly equivalent to the $\epsilon=0$ situation but shifted upward
slightly; for $\alpha\approx 2$ and above, where the outcome distribution is bifurcated,
the slight shift in the distribution's center causes positive outcomes
to be more likely than negative outcomes.

However, between these two outcomes lies a range of $\alpha$ where the $\epsilon=0$ case 
would have led to a bifurcation, but the lower tail of the distribution is suppressed
because the nobody-acts equilibrium is not feasible. In this range, we have an asymmetric
but unimodal distribution.

Figure \ref{ksplot} plots normalized kurtosis for each $\alpha$'s distribution. 
The neighborhood of $\alpha\approx 1.3$ is again salient, because the normalized kurtosis
in that range is an order of magnitude larger than three. The model's exceptional success
in generating a leptokurtic outcome makes the plot's vertical scale rather large, so it may be difficult to
discern that the kurtosis up to the peak is three, and after the peak is
one, as in the $\epsilon=0$ case.

The bottom plot of  Figure \ref{ksplot} shows that normalized skew
follows the same story relative to $\alpha$ as did kurtosis: it spikes
around 1.3, where the distribution of equilibrium percent acting has heavily negative skew.

Thus, given a realistic value of $\epsilon$ (i.e., anything but exactly zero), and a
value of $\alpha$ that is not too small to be equivalent to the private preferences
case and not too large to be equivalent to the full herding case, the distribution of
outcomes is unimodal, leptokurtic, and has a negative skew.

\section{Conclusion} \label{endsec}
There are several explanations for why rational agents would choose to
emulate others, all of which advise that a utility function meant to
describe a trader in the finance markets should include a term for the
desire to emulate others. 

Meanwhile, we know that equity return distributions show certain consistent 
deviations from the Normal distribution implied by na\"ive application
of a Central Limit Theorem. Adding a term for the emulation of others to
individual utilities produces aggregate outcome distributions that show
these same deviations from Normal: extreme outcomes happen more often,
and do so asymmetrically. 

However, the story is not quite as simple as saying that people tend to
imitate others. The type of distribution observed in equity returns appears in a
middle-ground between two extreme types of utility function.
With $\alpha$ small, the distribution of cutoffs is more-or-less that of
a situation of purely private utility. With $\alpha$ large, the
distribution follows the story of agents that simply follow the herd.
But between these two situations, there is a transition range where 
the distribution of cutoffs has the desired characteristics of being
unimodal, having large kurtosis, and skew toward zero. Thus, the model 
explains this type of distribution via an interplay between 
private and emulative utility.

This paper has shown that peer effects can generate leptokurtic outcomes under certain
conditions. This creates the possibility that an observed leptokurtic distribution 
can be explained by peer effects. For example, \citet{jones:budget} found leptokurtic
outcomes in Congressional actions such as budget allocations; I suggest in this paper
that a model of Congressional representatives who emulate each other can generate such
an outcome distribution. When outcomes have a blockbuster/flop bimodality, 
there is little doubt that peer effects are at play, but the model here
shows that even more subtle outcomes, with unimodal distributions but fat
tails, may also be the result of agents who gain direct or indirect utility from emulating each other.

\bibliographystyle{plainnat}
\bibliography{convivial}
\end{document}